\documentclass[12pt]{article}

\newcommand{\sect}[1]{\setcounter{equation}{0}\section{#1}}
\renewcommand{\theequation}{\arabic{section}.\arabic{equation}}

\def\be{\begin{equation}}
\def\ee{\end{equation}}
\def\ba{\begin{eqnarray}}
\def\ea{\end{eqnarray}}
\def\scri{{\cal I}}
\def\trho{\tilde{\rho}}

\title{{\bf Higher dimensional black holes and supersymmetry}}

\author{{\bf Harvey S. Reall} \\ Physics Department, 
Queen Mary, University of London \\ Mile End Road, London E1 4NS, 
United Kingdom}

\date{November 29 2002 \\ (Revised 13 December 2006) \\ 
 Preprint QMUL-PH-02-20 \\ hep-th/0211290}

\begin{document}

\maketitle

\begin{abstract}

It has recently been shown that the uniqueness theorem for stationary
black holes cannot be extended to five dimensions. However, uniqueness
is an important assumption of the string theory black hole entropy
calculations. This paper partially justifies this assumption  
by proving two uniqueness theorems for supersymmetric black holes in five
dimensions. Some remarks concerning general properties of
non-supersymmetric higher dimensional black holes are made. 
It is conjectured that there exist new families of 
stationary higher dimensional black hole
solutions with fewer symmetries than any known solution.

\end{abstract}

\sect{Introduction}

One of the most impressive successes of string theory is a microscopic
derivation of the entropy of certain supersymmetric black
holes \cite{strominger:96}.
The idea is that a weakly coupled system of strings and branes
wrapped around some compact dimensions turns into a black hole in the
non-compact dimensions as the string coupling is increased. For fixed
asymptotic charges (mass, angular momenta and gauge charges), the
degeneracy of microstates can be calculated in the weakly coupled
description. Provided sufficient supersymmetry is preserved, this is
found to reproduce correctly the Bekenstein-Hawking entropy of the
black hole (at least for black holes much larger than the string
length).

These calculations were first performed for static supersymmetric
black holes in five dimensions \cite{strominger:96}. 
They were subsequently extended to static
supersymmetric holes in four dimensions
\cite{maldacena:96,johnson:96}, to rotating supersymmetric
holes in five dimensions \cite{breckenridge:97}
and to nearly supersymmetric generalizations
of all of these \cite{horowitz:96a, breckenridge:96, horowitz:96b}. 

A key assumption made in this work is that the relevant black hole
solutions are uniquely specified by their asymptotic charges. If this
turned out to be untrue, i.e., if there existed distinct supersymmetric black
hole solutions with the same asymptotic charges, then there would be
a problem with the conventional interpretation of the entropy 
calculations. The problem would be in identifying which sets of
microstates should correspond to each black hole, as would be
necessary in order to compute their respective entropies. The black holes would
be distinguished macroscopically by their differing gravitational
fields. However, there is no gravitational field present in the weakly
coupled description used for the entropy calculations. Hence, at weak
coupling, there would be no way of telling which microstates should correspond to each black
hole. The distinction between different sets of microstates would only 
become apparent as the microscopic description became strongly coupled.  

Given the importance of this assumption, one might ask how it was
originally motivated. It seems that the only evidence in its favour
is the existence of the black hole uniqueness theorems
in four dimensions. These establish that stationary
four dimensional black holes are indeed uniquely specified by
their asymptotic charges, at least in Einstein-Maxwell theory. The
uniqueness theorems assume a non-degenerate event horizon and
therefore do not apply to supersymmetric black holes. Nevertheless, it
would be very surprising if supersymmetric black holes turned out to
be non-unique in four dimensions.\footnote{This paper will only
discuss spacetimes containing a single black hole. Otherwise
multi-black hole solutions \cite{hartle:72} 
would be an example of non-uniqueness.}

In five dimensions, the only evidence for the uniqueness
assumption seems to have been that higher dimensional black holes
appeared to have very similar properties to four dimensional ones.
However, this is not really evidence at all because
all known higher dimensional black hole solutions were derived 
using ans\"atze based on simple generalizations of four 
dimensional black hole solutions, or were related by dualities to
solutions based on such ans\"atze. Therefore it is not very surprising
that the known higher dimensional black holes
had similar properties to four dimensional ones.

The situation has changed with the recent discovery of a class of five
dimensional vacuum black holes that are completely unlike anything
encountered in four dimensions -- ``black rings'' \cite{emparan:02}. 
These are stationary black holes with event horizons of topology $S^1 \times
S^2$. They can be regarded as rotating loops of black string, with the
centrifugal force balancing the tendency of the ring to collapse under
gravity. The existence of black rings implies that the uniqueness theorem for 
stationary black holes does not extend to five dimensions. This is 
because black rings can carry the same asymptotic charges as the black 
holes of spherical topology discovered by Myers and Perry \cite{myers:86}.

In the first part of this paper, 
it will be suggested that there should be many more
exotic black hole solutions in higher dimensions. Examining
the steps that go into proving the uniqueness theorems in four
dimensions suggests that a {\it general} stationary asymptotically
flat black hole in
higher dimensions should admit only two commuting Killing vector
fields. However, all {\it known} higher dimensional black hole
solutions have more symmetry. So there may exist large families of
higher dimensional black hole solutions in addition to the known
ones. This would imply that black hole uniqueness would always be
badly violated in higher dimensions\footnote{
It is tempting to conjecture that adding the requirement of 
{\it stability} would guarantee uniqueness \cite{kol:02}, 
but there is no evidence for this since stability of
higher dimensional black holes has never been studied.}, 
and emphasizes the importance of
justifying the uniqueness assumption for supersymmetric black holes.

A uniqueness theorem {\it has} been proved for non-degenerate 
higher dimensional {\it static} black holes 
in Einstein-Maxwell \cite{gibbons:02a} and Einstein-Maxwell-dilaton
theory \cite{gibbons:02b}, and for Einstein gravity coupled to a
$\sigma$-model \cite{rogatko:02}.
The uniqueness assumption for static supersymmetric black
holes in higher dimensions therefore seems plausible. 

For rotating holes, it is not at all clear whether this assumption is
correct. Rotating supersymmetric black holes seem to exist only in
five dimensions -- the first example was found by Breckenridge,
Myers, Peet and Vafa (BMPV) \cite{breckenridge:97}.
It seems rather likely that charged black ring solutions should also
exist, and if these had a regular supersymmetric limit then uniqueness
of supersymmetric rotating black holes might be violated. Also, if
there do exist higher dimensional black holes with fewer symmetries
than any known solution 
then why not supersymmetric black holes with fewer symmetries than BMPV?
It is clearly desirable to know whether this happens 
and, if not, whether a uniqueness theorem for supersymmetric
black holes can be proved.

The main goal of this paper is to provide the first examples of 
uniqueness theorems for supersymmetric black holes, 
and thereby provide some justification for the
uniqueness assumption made in the black hole
entropy calculations. This is therefore a check on the consistency of
the entropy calculations that can be performed at the level of
classical supergravity.

The supergravity theory that will be considered
is minimal $D=5$ supergravity \cite{cremmer:81}
because it is the simplest theory
in which black hole uniqueness is known to be violated (the theory
admits black ring solutions). Furthermore, the BMPV supersymmetric
rotating black hole solution can be embedded in this
theory \cite{gauntlett:99}. In fact, this theory is sufficiently
simple that it is  possible to find {\it all}
supersymmetric solutions \cite{gauntlett:02}. Previously, the only
theories for which this had been done were minimal $N=2$, $D=4$
supergravity \cite{tod:83} and some simple $D=4$ generalizations
\cite{tod:95}. 

The general supersymmetric solution obtained in \cite{gauntlett:02} is
sufficiently complicated that it is far from obvious which solutions
correspond to black holes. In fact, the solution given in 
\cite{gauntlett:02} is only valid away from any horizons that may be
present in the spacetime. In this paper, it will be shown how a local
analysis of the constraints imposed by supersymmetry in the
neighbourhood of the horizon can be used to overcome this problem. The
following theorem will be proved:

\medskip

{\bf Theorem 1.} Any supersymmetric solution with a (spatially) compact horizon
has a near-horizon geometry that is locally isometric to one of the
following maximally supersymmetric solutions: flat space,
$AdS_3 \times S^2$, or the near-horizon geometry of the BMPV
solution (of which $AdS_2 \times S^3$ is a special case). 
The geometry of the horizon in these three cases is the
standard metric on $T^3$, $S^1 \times S^2$ or (a quotient of) a
squashed $S^3$, respectively. 

\medskip

Obviously it is desirable to extend this near-horizon result to a global uniqueness theorem. We shall see that this can be done for the BMPV black hole subject to one additional assumption. Supersymmetry guarantees the existence of a globally defined Killing vector field which is everywhere timelike or null \cite{gibbons:93}. The assumption is that this Killing field has no null orbits outside the black hole horizon:

\medskip

{\bf Theorem 2.} The only asymptotically flat supersymmetric black hole solution with near-horizon geometry locally isometric to the near-horizon BMPV solution and with the supersymmetric Killing vector field timelike everywhere outside the horizon is the BMPV black hole.

\medskip

These theorems constitute a uniqueness theorem for supersymmetric
black holes whose near-horizon geometry is not flat space or $AdS_3
\times S^2$. Note that this is much stronger than simply proving 
uniqueness for given horizon topology -- any supersymmetric black hole
other than BMPV must have an event horizon with {\it geometry} $T^3$ or
$S^1 \times S^2$. Of course, the latter possibility would describe 
a supersymmetric black ring so the above theorems do not exclude the
existence of such solutions. It will be interesting to see whether
they exist, or whether the theorems can be strengthened to eliminate
these exceptional cases.

This paper is organized as follows. Section 2 discusses general
properties of higher dimensional black holes. Section 3 contains the
uniqueness theorem. There is one appendix dealing with a special case
that arises in the analysis.

\sect{Higher dimensional black holes}

\subsection{Black holes with fewer symmetries}

All known stationary $D$-dimensional black hole solutions have at
least $[(D+1)/2]$ commuting isometries. The purpose of this section is
to point out that this seems to be ``too many'', i.e., in general one
would expect fewer symmetries. Before explaining this, it is helpful
to recall what happens for the analagous case of black string
solutions.

Consider the uniform black string solution of the five dimensional
vacuum Einstein equations. The
metric is the product of the four dimensional Schwarzschild solution
with a flat direction, so there are three commuting Killing vector
fields, corresponding to time translations, rotations and spatial
translations. If the string is compactified on a circle of asymptotic
radius $L$ then one can define a dimensionless parameter
\be
 \eta = \frac{G M}{L^2},
\ee 
where $M$ is the mass of the string. There is a particular value 
$\eta=\eta_c$ for which the uniform string solution admits a static
zero-mode that breaks the translational symmetry \cite{gregory:88}. 
This led to the conjecture \cite{gregory:88,horowitz:01} that exact 
static black string solutions without translational symmetry should also exist. 
There is good perturbative \cite{gubser:01} and numerical
\cite{wiseman:02,wiseman:02b} evidence that this is indeed the
case, but the solutions are not known analytically.\footnote{
See \cite{harmark:02,desmet:02} for attempts to construct such
solutions analytically.}
These solutions have only {\it two} commuting Killing vector fields, which
is one fewer than for the solutions that are known analytically. 

To understand why there might also exist stationary black {\it holes} with fewer
symmetries than any known solution, it is
worth reviewing the steps that go into proving the uniqueness
theorem for four dimensional black holes, and asking which steps can be
generalized to higher dimensions. For simplicity, only vacuum black
holes will be considered, although similar remarks should apply to
non-degenerate charged black holes. It is probably also worth emphasizing that only
asymptotically flat black holes will be considered in this paper.

The first step is the proof that the event horizon of a stationary
black hole must have $S^2$ topology \cite{hawking:72,hawking:73}. 
This relies on the Gauss-Bonnet theorem applied to the
(two dimensional) horizon and therefore does not generalize to higher
dimensions. An alternative proof in four dimensions 
is based on the notion of ``topological censorship''
\cite{friedman:93}. Consider a spacelike slice $\Sigma$ that
intersects the future event horizon and let $H$ denote the
intersection. Topological censorship requires that $\Sigma$ be simply
connected. Note that $\Sigma$ has two boundaries, namely $H$ and the
sphere at spatial infinity. Hence topological censorship requires
that $H$ be cobordant to a sphere via a simply connected
cobordism. For a stationary black hole, this can be shown to imply
that $H$ is a sphere \cite{chrusciel:94b}.

Topological censorship is also valid for $D>4$ but it is much
less restrictive. First, if $D>4$ and there exists a cobordism from $H$ to the
sphere then there also exists a simply connected
cobordism.\footnote{See \cite{hartnoll:03} for a recent review of
this, and other results from cobordism theory, with references
to the original literature.} Secondly, a
cobordism from $H$ to the sphere exists if, and only if, $H$ has
vanishing Pontrjagin and Stiefel-Whitney numbers. For $D=5$, $H$ is an oriented
3-manifold and hence automatically has vanishing Pontrjagin and
Stiefel-Whitney numbers so topological censorship 
does not restrict the topology of the event horizon for
$D=5$ black holes \cite{galloway:01}. For $D=6$, $H$ is a 4-manifold
and topological censorship excludes, for example, $H=CP^2$ because it
has non-vanishing Pontrjagin and Stiefel-Whitney numbers.

In summary, there are very few useful restrictions on the topology
of the event horizon of a general stationary black hole in higher
dimensions. However, black rings are the only known example of
stationary black holes with non-spherical horizons.

The next step is the four dimensional uniqueness 
proof that a stationary black hole must either be
static or have an ergoregion \cite{hawking:73}. This theorem
is straightforward to extend to higher dimensions. In the static case,
it can then be shown that the only solution is the Schwarzschild
solution \cite{israel:67,bunting:87}, and this theorem has recently been extended to
higher dimensions \cite{gibbons:02a}. A simple corollary is that a
static higher dimensional black hole must have a spherical horizon.

The possibility of an ergoregion disjoint from the event horizon was
excluded in \cite{sudarsky:90} for four dimensional black holes. This
proof relies on a technical theorem concerning maximal
hypersurfaces \cite{chrusciel:94}; 
it will be assumed here that it can be generalized to
higher dimensions. This implies that the stationary Killing vector field of a
stationary, non-static, higher dimensional black hole is spacelike on
the event horizon.

In four dimensions, it can be argued \cite{hawking:72, hawking:73} 
that the tangent vector to the
null geodesic generators of the event horizon can be extended to give a
Killing vector field $\xi$ of the full spacetime, which commutes with the
stationary Killing vector field. The latter cannot be equal to $\xi$
since it is spacelike on the horizon. One can therefore write (after
appropriately scaling $\xi$)
\be
 \xi = \frac{\partial}{\partial t} + \Omega \frac{\partial}{\partial \phi},
\ee
with $\partial/\partial \phi$ spacelike and Killing. It seems likely that this
theorem could be extended to higher dimensions although, since the
topology of the horizon is not known, the geometrical interpretation
of $\phi$ is not clear. Roughly speaking, this Killing vector should
correspond to a symmetry in the direction of rotation.

In four dimensions, the existence of two commuting Killing vector
fields implies that the metric has to take a fairly simple form, and it
can then be argued that any such solution to the Einstein equations
is uniquely determined by its mass and angular momentum
\cite{carter:71,robinson:75} and must belong to the Kerr family of
solutions. In higher dimensions, two Killing vector fields is not
enough symmetry to write the metric in a useful form, and the existence of
black rings shows that uniqueness should not be expected even when
more symmetry is present.

These general arguments suggest that all stationary higher dimensional
black holes must have two commuting symmetries. However, no known
higher dimensional black hole solution has {\it only} two commuting
symmetries. This suggests that higher dimensional black holes may be
similar to black strings in the sense that there may exist
undiscovered stationary solutions with fewer symmetries than the presently known solutions. 
More precisely,

\medskip

{\bf Conjecture.} There exist stationary, asymptotically flat black
hole solutions of the $D>4$ dimensional vacuum Einstein equations that
admit exactly two commuting Killing vector fields.

\medskip

These solutions would have to be non-static (because of the uniqueness
theorem for static black holes \cite{gibbons:02a}). 
If such solutions do exist then it seems unlikely that
the Schwarzschild solution would be recovered as a limit.
This would imply that such solutions must have an angular
momentum that is bounded below (in terms of their mass), just as
occurs for black rings. 

If the above conjecture is correct then higher dimensional black holes
would exhibit similar behaviour to black strings. There would be known
solutions with lots of symmetry and new solutions with less
symmetry. It is tempting to push this analogy further. Consider the
case of five dimensions with a single non-vanishing angular
momentum. Define a dimensionless parameter $\eta$ by
\be
 \eta \equiv \frac{27 \pi J^2}{32 G M^3},
\ee
where $J$ and $M$ are the angular momentum and mass of a black
hole. The known solutions are the Myers-Perry solutions
\cite{myers:86} (which exist for $\eta<1$) and black rings
\cite{emparan:02} ($\eta>\eta_* \approx 0.84$). These
solutions have three commuting Killing vector fields
$\partial/\partial t$, $\partial/\partial \phi$ and $\partial/\partial
\psi$ where $\phi$ is the direction of rotation. The above conjecture
suggests looking for new solutions without symmetry in the $\psi$
direction. The analogy with black strings suggests that there might be some critical
value $\eta=\eta_c$ for which the Myers-Perry solution (or black ring)
admits a stationary zero-mode that breaks the symmetry in the $\psi$ direction. 
Finding such a mode would therefore be evidence in favour
of the above conjecture.\footnote{Examining
perturbations of Myers-Perry solutions would also be of interest in
view of the conjecture \cite{emparan:02} that a five dimensional
Myers-Perry black hole with a single non-vanishing angular momentum
is classically unstable for $\eta$ close to $1$.} However, the
absence of such a mode would not rule out the existence of new
solutions. For example, the topology of the new solutions might differ
from that of the Myers-Perry solutions and black rings, in which case
they would not be seen in perturbation theory about the known
solutions.

\subsection{Magnetic rings}

The existence of black rings implies that stationary black holes in
five dimensions are not uniquely specified by their asymptotic
charges. If the above conjecture is correct then there exist further
black hole solutions, and therefore black hole uniqueness is more
severely violated. It is clearly desirable to know {\it how many} stationary
higher dimensional black hole solutions have a given set of asymptotic charges. Are
there finitely many or infinitely many? The purpose of the present
subsection is to suggest that there may be a {\it continuous} infinity
of solutions with a given set of asymptotic charges.

The black ring solutions obtained in \cite{emparan:02} are solutions
of the vacuum Einstein equations in five dimensions. It is interesting
to ask whether electromagnetic generalizations exist. Consider
Einstein-Maxwell theory in five dimensions, possibly with a
Chern-Simons term. This theory
admits two types of static black string solution: electric and
magnetic. The electric solution becomes nakedly singular in the
extremal limit. The extremal solution is best viewed as a smeared
distribution of black holes. The magnetic solution has a regular
extremal limit: this is the supersymmetric black string of minimal
$N=1$, $D=5$ supergravity.

Black rings can be regarded as rotating loops of black string. 
Consider a rotating loop of magnetic black string. If such a solution exists
then it would have vanishing electric charge.\footnote{Hence it could
not saturate the Bogomolnyi bound appropriate to an asymptotically
flat spacetime \cite{gibbons:93} and therefore
would not be supersymmetric.} 
The magnetic charge of a localized configuration must vanish
in four spatial dimensions \cite{dowker:96}. Therefore the only asymptotic
charges that would be carried by such a solution are its mass and
angular momentum. However, the solution would presumably be
characterized by a third parameter $\alpha$ measuring the strength of the
magnetic field. Therefore, if magnetic black rings exist, then they
would be an example of a continuous family (labelled by $\alpha$) 
of solutions with the same asymptotic charges.

\sect{Uniqueness theorems}

\subsection{Introduction}

The above considerations highlight how little is known about general properties
of higher dimensional stationary black holes, and suggest that such black holes
are highly non-unique, if non-static. This casts doubt on the
uniqueness assumption that underlies the entropy calculations for
supersymmetric rotating black holes \cite{breckenridge:97}. A
uniqueness theorem is required in order to justify this
assumption. In this section, the theorems stated in the introduction
will be proved.

Proving uniqueness theorems for supersymmetric black holes is much
easier than, say, attempting to generalize the known black hole
uniqueness theorems to include degenerate horizons. This is because the
existence of a globally defined super-covariantly constant spinor
highly constrains the form of the spacetime. In fact, for minimal
$N=2$, $D=4$ supergravity, it fully determines the local form of the
metric \cite{tod:83}. For minimal $D=5$ supergravity, a simple
algorithm can be given for the construction of all supersymmetric
solutions \cite{gauntlett:02}. This will be reviewed in subsection
\ref{subsec:sugra}.

The method of \cite{gauntlett:02} yields the general supersymmetric
solution in a coordinate system that does not cover any event horizons
in the spacetime. Therefore, the first step in the uniqueness proof is
to introduce a coordinate system valid in the neighbourhood of a
Killing horizon (subsection \ref{subsec:coords}), and to repeat some 
of the analysis of \cite{gauntlett:02} in these coordinates
(subsection \ref{subsec:nearhorsusy}). It turns out that this fully
determines the local form of the near-horizon geometry (subsections
\ref{subsec:special} and \ref{subsec:nearhor}), thereby proving Theorem 1. 
Theorem 2 is proved by showing that  
the local form of the near-horizon geometry, together with asymptotic flatness, is
sufficient to select a unique solution from the general solution of
\cite{gauntlett:02}, which must therefore be the known BMPV solution
(subsection \ref{subsec:global}).  

\subsection{Minimal five dimensional supergravity}

\label{subsec:sugra}

Minimal $N=1$, $D=5$ supergravity was constructed in
\cite{cremmer:81}. The bosonic sector has action\footnote{
Conventions: the metric has positive signature, curvature is defined so that de
Sitter space has positive Ricci scalar. Curved indices are denoted by
$\mu,\nu\ldots$ and tangent space indices by $\alpha,\beta,\ldots$.}
\be 
 \label{eqn:action}
 S = \frac{1}{4\pi G} \int \left( \frac{1}{4} R * 1 - \frac{1}{2} {}F
 \wedge *{}F - \frac{2}{3\sqrt{3}} {}F \wedge {}F \wedge A \right).
\ee
All purely bosonic supersymmetric solutions of this theory were
obtained in \cite{gauntlett:02} as follows.
Starting from a commuting super-covariantly constant (Dirac) spinor
$\epsilon$, one can construct a real scalar field $f$, a real vector field
$V$ and three real two-form fields $X^{(i)}$:\footnote{The precise
definition of these objects is given in \cite{gauntlett:02} in terms
of symplectic-Majorana spinors. Converting to Dirac spinors may
introduce numerical factors, which have not been calculated here.} 
\ba
 f &\sim & i \bar{\epsilon} \epsilon, \qquad V^{\alpha} \sim
 \bar{\epsilon} \gamma^{\alpha} \epsilon, \nonumber \\
 \left(X^{(1)}+iX^{(2)}\right)_{\alpha\beta} 
 &\sim  &\epsilon^T C \gamma_{\alpha\beta}
 \epsilon, \quad X^{(3)}_{\alpha\beta} \sim \bar{\epsilon}
\gamma_{\alpha\beta} \epsilon.
\ea
Fierz identities imply various algebraic identities between these
quantities, for example
\be
\label{eqn:fV}
 f^2 = -V^2,
\ee
\be
\label{eqn:Xcond1}
 i_V X^{(i)} = 0,
\ee
\be
\label{eqn:Xcond2}
 i_V * X^{(i)} = - f X^{(i)},
\ee
\be
 X^{(i)}_{\gamma \alpha} X^{(j)\gamma}{}_{\beta} = \delta_{ij} \left(f^2
 \eta_{\alpha\beta} +V_{\alpha} V_{\beta} \right) - f \epsilon_{ijk}
 X^{(k)}_{\alpha \beta},
\ee
where $\epsilon_{123} = +1$ and,
for a $p$-form $A$ and a vector $Y$, $i_Y A$ denotes the
$(p-1)$-form obtained by contracting $Y$ with the first index of $A$.
Equation \ref{eqn:fV} implies that the
vector field $V$ is timelike, null or zero. Since $V_0 \sim
\epsilon^\dagger \epsilon$, the latter possibility occurs if, and only
if, $\epsilon$ vanishes. Since $\epsilon$ is super-covariantly constant, 
the above quantities must also satisfy certain differential constraints 
\cite{gauntlett:02}:
\be
\label{eqn:df}
 df = -\frac{2}{\sqrt{3}} i_V F,
\ee
\be
 D_{(\alpha } V_{\beta)} = 0,
\ee
\be
 dV = -\frac{4}{\sqrt{3}} fF - \frac{2}{\sqrt{3}} * \left( F \wedge V \right),
\ee
and
\be 
\label{eqn:gradX}
 D_\alpha X^{(i)}_{\beta\gamma} = \frac{1}{\sqrt{3}} \left[ 2
 {}F_{\alpha}{}^{\delta} \left( *X^{(i)} \right)_{\delta\beta\gamma} -2
 {}F_{[\beta}{}^{\delta} \left( *X^{(i)} \right)_{\gamma] \alpha
 \delta} + \eta_{\alpha [\beta} {}F^{\delta \epsilon} \left( *X^{(i)}
 \right)_{\gamma] \delta\epsilon} \right],
\ee 
which implies 
\be
\label{eqn:dX}
 d X^{(i)} = 0.
\ee
These equations imply that $V$ is a Killing vector field that
preserves the field strength (i.e. ${\cal L}_V F = 0$ where ${\cal L}$
denotes the Lie derivative), i.e., $V$ generates a symmetry of the
full solution.

If $p$ is a point at which $V$ vanishes then consider a timelike
geodesic through $p$. Let $U$ denote the tangent vector to this
geodesic. $V$ is a Killing vector field so $V \cdot U$ is conserved
along the geodesic, and must therefore vanish because it vanishes at
$p$. Therefore $U$ and $V$ are orthogonal along the geodesic. However,
$U$ is timelike and $V$ is non-spacelike so this implies that $V$ must
vanish everywhere along the geodesic, and therefore so must $\epsilon$.
This applies to all timelike geodesics through $p$. Hence $\epsilon$ 
vanishes in open regions to the future and past of $p$. By
analyticity, $\epsilon$ must then vanish everywhere, which contradicts
the assumption that the spacetime admits a super-covariantly constant
spinor. Hence there cannot exist any point in the spacetime at which
$V$ or $\epsilon$ vanishes.

Either $f$ vanishes throughout the spacetime or there is some point
$p$ at which $f \ne 0$. These will be referred to as the
``null case'' and ``timelike case'' respectively. In the null case,
$V$ is a globally defined null Killing vector field $V$. 
In fact the general solution in this case is a plane-fronted wave 
\cite{gauntlett:02}. Special cases of this general solution include
the magnetic black string solution \cite{gibbons:94} and its near
horizon geometry, $AdS_3 \times S^2$. The existence of a globally
defined null Killing vector field implies that these solutions cannot
describe black holes.

In the timelike case, by continuity, there is some topologically
trivial neighbourhood ${\cal U}$ of $p$ in which $f \ne 0$. 
Therefore $V$ is a timelike Killing vector field in ${\cal U}$. 
It will be assumed that $f>0$ without loss of generality 
\cite{gauntlett:02}. Coordinates can be introduced so that 
the metric in ${\cal U}$ can be written \cite{gauntlett:02}
\be
\label{eqn:metricsol}
 ds^2 = -f^2 \left( dt + \omega \right) + f^{-1} ds_4^2,
\ee
where $V = \partial/\partial t$ and $ds_4^2$ is the metric on a four
dimensional Riemannian ``base space'' orthogonal to the orbits of
$V$. Note that all metric components must be independent of $t$.
$\omega$ is a $1$-form that is defined by the equations
\be
\label{eqn:omegadef}
 i_V \omega =0, \qquad d \omega = -d\left(f^{-2} V \right).
\ee
This determines $\omega$ up to a gradient, which reflects the freedom
to choose the $t=0$ hypersurface. Supersymmetry requires that the base
space be hyper-K\"ahler, with $X^{(i)}$ the three complex structures
and a volume form $\eta_4$ chosen so that these are anti-self-dual. 
This volume form is related to the volume form $\eta$ on the five
dimensional spacetime by
\be
\label{eqn:eta4}
 \eta_4 = f i_V \eta.
\ee
$d\omega$ can be regarded as a $2$-form on the base space and
can therefore be decomposed into self-dual and anti-self-dual parts
with respect to the base space:
\be
 f d\omega = G^+ + G^-.
\ee
It is then possible to solve for the field strength \cite{gauntlett:02}:
\be
\label{eqn:Fsol}
 F = -\frac{\sqrt{3}}{2} d \left[ f^{-1} V \right] - \frac{1}{\sqrt{3}} G^+.
\ee
The Bianchi identity for $F$ yields
\be
\label{eqn:dGplus}
 dG^+ = 0,
\ee
and the equation of motion for $F$ gives
\be
\label{eqn:delf}
 \Delta f^{-1}=\frac{4}{9}(G^{+})^2, 
\ee 
where $\Delta$ is the Laplacian associated with the base space metric
and 
\be
 (G^{+})^2\equiv \frac{1}{2} (G^{+})_{mn}(G^{+}){^{mn}},
\ee
where $m,n$ are indices on the base space, raised with the base
space metric. The above equations guarantee that equations
\ref{eqn:metricsol} and \ref{eqn:Fsol} yield a supersymmetric solution
of the supergravity theory \cite{gauntlett:02}. 

Any supersymmetric black hole solution must belong to the
timelike class. Therefore the full black hole spacetime is determined
by analytic continuation of a solution of the above form. The only known
supersymmetric black hole solution of this theory is the BMPV black
hole \cite{breckenridge:97,gauntlett:99}, which has
base space $R^4$, with metric
\be
 ds_4^2 = d\rho^2 + \frac{\rho^2}{4} \left[ \left(\sigma_R^1\right)^2 +
   \left(\sigma_R^2 \right)^2 + \left( \sigma_R^3 \right)^2 \right],
\ee
where $\sigma_R^i$ are left invariant $1$-forms on $SU(2)$ -- see
\cite{gauntlett:02} for details. The solution has
\be
 \omega = \frac{j}{2\rho^2} \sigma_R^3,
\ee
which implies $G^+ = 0$. The solution for $f$ is
\be
 f^{-1} = 1 + \frac{\mu}{\rho^2}.
\ee
The global properties of this solutions were investigated in detail in
\cite{gauntlett:99, gibbons:99}. The solution describes a black hole 
provided $j^2 < \mu^3$. If this bound is violated then it instead
describes a regular spacetime with naked closed causal curves
\cite{gibbons:99} and the microscopic description becomes non-unitary
\cite{herdeiro:00}. There exists evidence \cite{gibbons:99, jarv:02}
that it is physically impossible to add angular momentum to the black
hole and violate the above bound.

\subsection{Introduction of coordinates}

\label{subsec:coords}

The coordinate system introduced above is only valid locally, and
does not cover regions in which
$f$ vanishes, for example the event horizon of a black hole. In this
section, a new set of coordinate will be introduced that do
cover such a horizon. However, before doing this, it is necessary to
argue that, for a supersymmetric black hole solution, the Killing
vector field $V$ has the usual properties associated with the
stationary Killing vector field of an equilibrium black hole spacetime. 

First, consider the possibility that $V$ becomes null at some point
$p$ outside the black hole. Equation \ref{eqn:df} implies $V \cdot
\partial f=0$, so $f$ is constant along the orbits of $V$. Since $f$
vanishes at $p$, it must vanish along the orbit through $p$. Hence $V$
is null on this orbit. However, one would not expect a spacetime
describing the rest frame of a {\it single} black hole to admit a
Killing vector field with a null orbit {\it outside} the black
hole. This is because such an orbit would correspond to an observer
moving at the speed of light for whom the gravitational field would appear
unchanging. We would expect that this is only possible on the event horizon.
Therefore, we assume that $V$ is timelike everywhere outside the black hole,
i.e. it will be assumed that $f>0$ everywhere outside the black hole. 

Now consider the behaviour of $V$ at infinity. If $V$ were to vanish 
at some point $p$ on $\scri^+$ then consider an affinely
parametrized outgoing null geodesic with an endpoint at $p$. Let $k$
denote the tangent vector to this geodesic. $V \cdot k$ is constant
along the geodesic and vanishes at infinity. Hence $V \cdot k$
vanishes everywhere on the geodesic. Therefore $V$ must be proportional to
$k$, and therefore null, along this geodesic. But the argument above
excludes the possibility of $V$ being null outside the black
hole. Hence $V$ cannot vanish on $\scri^+$. Similarly, $V$ cannot
vanish on $\scri^-$. 

If $V$ were to become null at some point $p$ on $\scri^+$ then it is
easy to see that $V$ must be everywhere tangent to the null geodesic
generator of $\scri^+$ through $p$. Once again, such a null symmetry
would not be expected of a spacetime describing a black hole at rest
so $V$ cannot be null anywhere on $\scri^+$. Similarly, $V$ cannot be
null anywhere on $\scri^-$. 

These considerations establish that $V$ must be timelike everywhere
outside the black hole and also on $\scri^{\pm}$. If $f$ were to
diverge anywhere on $\scri^{\pm}$ then $V$ would be behaving as a
boost symmetry, which is not expected for a black hole in its rest
frame. Hence $f$ must be non-zero and bounded on $\scri^{\pm}$. 

It will be assumed that the future event horizon ${\cal H}^+$ has a
single connected component. Since $V$ is an isometry, it must leave
this horizon invariant and must therefore be null on ${\cal
H}^+$. 

Let $\Sigma$ be a Cauchy surface for the exterior region of the black
hole such that $\Sigma$ has a boundary $H$ on the future event
horizon. A null Gaussian coordinate system can be set up in a
neighbourhood of $H$ as follows (see \cite{friedrich:99} for more details).
Introduce local coordinates $x^A$ ($A=1,2,3$) on $H$. Let $p$ be a point
on $H$ with coordinates $x^A$. Consider the future directed null
geodesic generator of ${\cal H}^+$ that passes through $p$, with
tangent vector $V$. The
coordinates of a point affine parameter distance $u$ 
from $p$ along this generator
will be defined to be $(u,x^A)$. This defines coordinates on a
neighbourhood ${\cal U}$ of $H$ in ${\cal H}^+$ with $V = \partial/\partial u$. 
Now let $n$ be the unique past directed null vector field defined on 
${\cal U}$ by $V\cdot n = 1$ and $n \cdot X = 0$ for all $X$
tangent to surfaces of constant $u$. Finally, consider the null
geodesic from a point $p \in {\cal U}$ with tangent $n$. Let the coordinates
of a point affine parameter distance $r$ along this geodesic be
$(u,r,x^A)$ where $(u,x^A)$ are the coordinate of $p$. 

It is easy to check that ${\cal L}_V n = 0$ on ${\cal H}^+$. Moreover,
$V$ is a Killing vector field and hence geodesics are mapped to
geodesics under the flow of $V$. Putting these facts together, under
the flow of $V$ through a parameter distance $\delta$, the point with
coordinates $(u,r,x^A)$ is mapped to the point with coordinates
$(u+\delta,r,x^A)$. Hence 
\be
 V = \partial/\partial u
\ee
everywhere, not just on the horizon.

Since $f$ vanishes at $r=0$, differentiability implies
\be
 f = r \Delta(r,x^A),
\ee
for some function $\Delta$ independent of $u$ (as $V \cdot \partial f
= 0$). The exterior of the black hole is $r>0$ so $\Delta$ must be
positive for $r>0$. $\partial/\partial x^A$ is tangent to surfaces of
constant $u$ in ${\cal H}^+$ and hence orthogonal to $V$ at
$r=0$. Therefore $g_{uA} = r h_A (r,x^B)$ for some functions $h_A$
independent of $u$ (as $V$ is Killing). The full metric must take the
form \cite{friedrich:99}
\be
\label{eqn:metric}
 ds^2 = -r^2 \Delta^2 du^2 + 2 du dr + 2 r h_A du dx^A + \gamma_{AB} dx^A dx^B,
\ee
where $\gamma_{AB}$ is a function of $r$ and $x^A$. It was argued
above that black holes must belong to the timelike family of
solutions, for which $\Delta > 0$ for $r>0$. However, the above line
element is clearly also valid in the neighbourhood of a Killing
horizon of $V$ in the {\it null} family, for which $\Delta \equiv
0$. Also note that the form of this metric guarantees the existence of
a regular near horizon geometry, defined by the limit $r = \epsilon
\tilde{r}$, $u = \tilde{u}/\epsilon$ and $\epsilon \rightarrow 0$.

\subsection{Supersymmetry near the horizon}

\label{subsec:nearhorsusy}

The next step in the proof is to examine the constraints imposed by
supersymmetry in the above coordinate system.

Using the above form for the metric, equations \ref{eqn:Xcond1} and
\ref{eqn:Xcond2} imply that the two forms can be written
\be
\label{eqn:Zdef}
 X^{(i)} = dr \wedge Z^{(i)} + r \left( h \wedge Z^{(i)} - \Delta *_3
 Z^{(i)} \right),
\ee
where $Z^{(i)} \equiv Z^{(i)}_A dx^A$, $h \equiv h_A dx^A$ and $*_3$
denotes the Hodge dual with respect to $\gamma_{AB}$. $X^{(i)}$ is
globally defined so $Z^{(i)}$ are well-defined in a neighbourhood of $H$.
The algebraic relations satisfied by $X^{(i)}$ imply
\be
 <Z^{(i)},Z^{(j)}> = \delta^{ij}, \qquad Z^{(i)} \wedge Z^{(j)} =
   \epsilon_{ijk} *_3 Z^{(k)},
\ee
where $<,>$ denotes the inner product defined by
$\gamma_{AB}$. Closure of $X^{(i)}$ (equation \ref{eqn:dX}) yields
\be
 \hat{d} Z^{(i)} = -\frac{1}{2} \partial_r (r\Delta) \epsilon_{ijk}
 Z^{(i)} \wedge Z^{(j)} + \partial_r (rh) \wedge Z^{(i)} - r\Delta
 \epsilon_{ijk} \partial_r Z^{(j)} \wedge Z^{(k)} + r h \wedge
 \partial_r Z^{(i)},
\ee
and
\be
\label{eqn:dh}
 *_3 \hat{d} h - \hat{d} \Delta - \Delta h + r \partial_r \Delta h -
 2r \Delta \partial_r h -r *_3 \left(h \wedge \partial_r h \right) - r
 \Delta^2 \epsilon_{ijk} Z^{(i)} < Z^{(j)}, \partial_r Z^{(k)} > = 0,
\ee
where, for a $p$-form $Y$ with only $A,B,C$ indices,
\be
  (\hat{d} Y)_{ABC\ldots} \equiv (p+1) \partial_{[A} Y_{BC \ldots ]}.
\ee
For $r>0$, $\omega$ can be defined as in equation \ref{eqn:omegadef}, 
giving
\be
 \omega = -\frac{dr}{r^2 \Delta^2} - \frac{1}{r\Delta^2} h,
\ee
where an arbitrary gradient can be absorbed by shifting the $u=0$ surface.
The definition of $G^+$ can be rewritten as
\be
 G^+ = \frac{1}{2} \left( f d \omega + i_V * d\omega \right).
\ee
Computing $G^+$ then gives 
\be
\label{eqn:Gplussol}
 G^+ = dr \wedge {\cal G} + r \left( h \wedge {\cal G} + \Delta *_3 {\cal G} \right),
\ee
where
\be
 {\cal G} = -\frac{3}{2r \Delta^2} \hat{d} \Delta +
 \frac{3}{2\Delta^2} \partial_r \Delta h - \frac{3}{2 \Delta}
 \partial_r h - \frac{1}{2} \epsilon_{ijk} Z^{(i)} <Z^{(j)},
 \partial_r Z^{(k)} >.
\ee
Equation \ref{eqn:dh} was used to eliminate $\hat{d}h$ from this
expression. Note the similarity between equations \ref{eqn:Zdef} and
\ref{eqn:Gplussol}: this structure is a consequence of (anti)-self-duality on
the base space. Equation \ref{eqn:Fsol} now gives
\ba
\label{eqn:Fsol2}
 F &=& \frac{\sqrt{3}}{2} \left[ -\partial_r (r\Delta) du \wedge dr -
 r du \wedge \hat{d} \Delta + \frac{1}{3} \epsilon_{ijk} dr \wedge
 Z^{(i)} < Z^{(j)}, \partial_r Z^{(k)} > - *_3 h \right. \nonumber \\
 &-& \left. r *_3 \partial_r h + \frac{r}{3} \epsilon_{ijk} \left(
 -2\Delta *_3 Z^{(i)} + h \wedge Z^{(i)} \right)< Z^{(j)}, \partial_r
 Z^{(k)} >  \right].
\ea
Note that this is well-defined at $r=0$, and has a well-defined
near-horizon limit, even though $G^+$ need not be regular at
$r=0$.

The $ABC$ component of the Bianchi identity for $F$ (or equivalently,
equation \ref{eqn:dGplus}) now yields
\be
\label{eqn:dstarh}
 \hat{d} *_3 h = {\cal O}(r).
\ee
Now equations \ref{eqn:dh} and \ref{eqn:dstarh} give
\be
 \hat{d} \Delta \wedge *_3 \hat{d} \Delta = \hat{d} \left[ \Delta
 \hat{d} h - \frac{1}{2} \Delta^2 *_3 h \right] + {\cal O} (r).
\ee
Integrating this equation over the compact $3$-manifold $H$ (which is
at $r=0$) then implies
\be
 \hat{d} \Delta = 0 \qquad \mbox{on $H$},
\ee
hence $\Delta$ is constant on the event horizon. Equations
\ref{eqn:dh} and \ref{eqn:dstarh} now yield
\be
\label{eqn:honH}
 \hat{d}h = \Delta *_3 h, \qquad \hat{d} *_3 h = 0 \qquad \mbox{on $H$}.
\ee
The $ArB$ component of equation \ref{eqn:gradX} gives
\be
\label{eqn:gradZ}
 \nabla_A Z^{(i)}_B = - \frac{1}{2} \Delta \left( *_3 Z^{(i)}
 \right)_{AB} + \gamma_{AB} <h,Z^{(i)}> - Z^{(i)}_A h_B + {\cal O}(r),
\ee
where $\nabla$ is the connection associated with $\gamma_{AB}$. Taking
another derivative and antisymmetrizing then yields an expression for
the Riemann tensor of $H$. From this, the Ricci tensor of $H$ is (using \ref{eqn:honH})
\be
\label{eqn:ricH1}
 R_{AB} = \left(\frac{\Delta^2}{2} + h^2 \right) \gamma_{AB} -
 \nabla_{(A} h_{B)} - h_A h_B,
\ee
where $h^2 = h_A h^A$, raising indices with $\gamma^{AB}$ on $H$.
Equations \ref{eqn:honH} imply
\be
\label{eqn:laph}
 \nabla^2 h_A = R_{AB} h^B - \Delta^2 h_A \qquad \mbox{on $H$}.
\ee
Now consider
\be
 I = \int_H \nabla_{(A} h_{B)} \nabla^{(A} h^{B)}.
\ee
Integrating by parts and using \ref{eqn:laph} and $\nabla_A h^A = 0$
(from \ref{eqn:honH}) gives
\be
 I = \int_H \left(\Delta^2 h^2 - 2 R_{AB} h^A h^B \right).
\ee
Finally, substituting in equation \ref{eqn:ricH1} and integrating by
parts yields $I=0$. Hence
\be
\label{eqn:hkilling}
 \nabla_{(A} h_{B)} = 0 \qquad \mbox{on $H$}.
\ee
Therefore, if non-zero, then $h$ is a Killing vector field on
$H$. Substituting into \ref{eqn:ricH1} gives the Ricci tensor of $H$
\be
\label{eqn:ric}
 R_{AB} = \left(\frac{\Delta^2}{2} + h^2 \right) \gamma_{AB} - h_A
 h_B.
\ee
This completely determines the curvature of $H$ because $H$ is a $3$-manifold.
Combining equations \ref{eqn:honH} and \ref{eqn:hkilling} gives
\be
\label{eqn:gradh}
 \nabla_A h_B = \frac{1}{2} \Delta \eta_{ABC} h^C,
\ee
where $\eta_{ABC}$ is the volume form of $H$. Note that this implies
that $h^2$ is constant on $H$. Furthermore, combining equations
\ref{eqn:gradZ} and \ref{eqn:gradh} gives
\be
 [h,Z^{(i)}] = 0 \qquad \mbox{on $H$}.
\ee

\subsection{A special case}

\label{subsec:special}

This subsection will consider the case in which $\Delta$ vanishes on
$H$. If this happens then, on $H$,
\be
\label{eqn:dzspecial}
 \hat{d} Z^{(i)} = h \wedge Z^{(i)},
\ee 
which implies that the $1$-forms $Z^{(i)}$ are hypersurface
orthogonal, i.e, there exist functions $z^i$ and $K^{(i)}$
defined on $H$ so that $Z^{(i)} = K^{(i)} dz^i$ (no summation on $i$).
Equation \ref{eqn:dzspecial} then requires that the functions
$K^{(i)}$ be proportional. The constants of proportionality can be
absorbed into $z^i$, so $K^{(i)} = K$ for $i=1,2,3$, i.e.,
\be
 \hat{d}Z^{(i)} = K dz^i.
\ee
Equation \ref{eqn:dzspecial} also implies
\be
 h = \hat{d} \log K.
\ee
The functions $z^i$ can be used as local coordinates on $H$, i.e.,
$\{x^A \} = \{z^i\}$. Orthonormality of $Z^{(i)}$ implies that the
metric on $H$ is conformally flat:
\be
 \gamma_{AB} dx^A dx^B = K^2 dz^i dz^i.
\ee
Equation \ref{eqn:gradh} says that $h$ is covariantly constant. In
these coordinates, this gives
\be
 K^{-1} \partial_i \partial_j K - 3 K^{-2} \partial_i K \partial_j K +
 K^{-2} \partial_k K \partial_k K \delta_{ij} = 0.
\ee
This equation was encountered in \cite{gauntlett:02}. By shifting the
origin and rescaling the coordinates $z^i$, the solutions can be
written
\be
 K = 1, \qquad {\rm or} \qquad K = \frac{L}{R},
\ee
where $R = \sqrt{z^i z^i}$. In the first case, this implies that the
metric near $H$ can be written
\be
 ds^2 = 2 dr du + \left(\delta_{ij} + {\cal O}(r) \right) dz^i dz^j +
 {\cal O}(r^4) du^2 + {\cal O}(r^2) du dz^i. 
\ee
The near horizon limit of this solution is locally isometric to 
flat space with vanishing gauge field. 
Globally it must differ by some discrete identifications
because $H$ is assumed compact. The metric on $H$ is flat so $H$ must
be some quotient of $R^3$ with its flat metric
\cite{giulini:94}, i.e., $R^3$ identified with respect to some
subgroup of its isometry group. However, these identifications have to
preserve the 1-forms $Z^{(i)}$, which implies that they must be
translations. So $H$ is a compact manifold obtained by identifying
$R^3$ with respect to certain translations, i.e., $H$ must be a 3-torus $T^3$. 

In the second case, it is convenient to use spherical
polar coordinates $\{x^A\} = (R,\theta,\phi)$. The solution can be written
\be
\label{eqn:ads3}
 ds^2 = 2 dr du - 2 \frac{r}{R} du dR + L^2 
\left( \frac{dR^2}{R^2} + d\theta^2 + \sin^2
 \theta d\phi^2 \right) +
{\cal O}(r^4) du^2 + {\cal O}(r^2) du dx^A + {\cal O}(r) dx^A dx^B.
\ee
The near horizon limit of such a solution is
locally isometric to the (maximally supersymmetric) $AdS_3 \times S^2$ 
solution (to see this, let $r=vR/L$ for some new coordinate $v$).
Globally, it must differ because $H$ is compact. The metric on $H$ can
be written
\be
 ds_3^2 = L^2 \left( dZ^2 + d\theta^2 + \sin^2 \theta d\phi^2 \right),
\ee
where $Z = \log R$ (note that $R$ must be bounded away from zero
because the 1-forms $Z^{(i)}$ are well-defined on $H$). Hence $H$ is
locally isometric to the standard metric on $R \times S^2$, which
presumably implies that $H$ can be obtained as a quotient of $R \times
S^2$. However, the only elements of the isometry group of $R \times
S^2$ which preserve the 1-forms $Z^{(i)}$ are translations $Z
\rightarrow Z+{\rm constant}$. Hence $H$ must be globally isometric to
the standard metric on $S^1 \times S^2$ with $Z \sim Z + l$ for some $l$. 

The $AdS_3 \times S^2$ solution arises as the near-horizon geometry 
of a momentum-carrying black {\it string} wrapped around a 
compact Kaluza-Klein direction\footnote{
If the string does not carry momentum then it cannot be identified to
yield a regular compact event horizon \cite{gibbons:94}.}. To fully
understand uniqueness of supersymmetric black holes it is necessary to
know whether $AdS_3 \times S^2$ can also arise as the near-horizon
geometry of a black {\it hole}. Such a solution would have horizon
topology $S^1 \times S^2$, i.e., it would be a supersymmetric black ring. 
Some restrictions on the geometry of the base space of such solutions (if
they exist) are obtained in the Appendix.

\subsection{Near horizon geometry}

\label{subsec:nearhor}

Now consider the case in which $\Delta >0$ on $H$. It will
be shown that the near horizon geometry must be locally 
isometric to that of the BMPV black hole. First define
a set of $1$-forms on $H$ by\footnote{All equations in this subsection
are evaluated on $H$, so hats will not be included on exterior derivatives.}
\be
 \sigma_L^i = \frac{\Delta^2 + h^2}{\Delta} Z^{(i)} + \frac{1}{\Delta}
 d \left(h^A Z^{(i)}_A \right).
\ee
Using equations \ref{eqn:gradZ} and \ref{eqn:gradh}, it can be shown
that these $1$-forms obey
\be
 d \sigma_L^i = -\frac{1}{2} \epsilon_{ijk} \sigma_L^j \wedge
 \sigma _L^k,
\ee
and
\be
 [h,\sigma_L^i] = 0.
\ee
The vector fields dual to these $1$-forms are
\be
 \xi^i_L = \frac{\Delta}{\Delta^2 + h^2} Z^{(i)} - \frac{1}{\Delta^2 +
   h^2} *_3 \left( h \wedge Z^{(i)} \right).
\ee
Equations \ref{eqn:gradZ} and \ref{eqn:gradh} imply that these are
Killing vector fields:
\be
 \nabla_{(A} \left(\xi^i_L \right)_{B)} = 0,
\ee
and satisfy the commutation relations of $SU(2)$:
\be
 [\xi_L^i,\xi_L^j] = \epsilon_{ijk} \xi_L^k.
\ee
Furthermore, these Killing vector fields commute with $h$:
\be
 [h,\xi_L^i] = 0.
\ee
These considerations show that, if $h \ne 0$ then $H$ admits four
globally defined Killing vector fields satisfying the commutation
relations of $SU(2) \times U(1)$. 

If $h \ne 0$ then local coordinates can now be introduced as
follows. Let $\hat{h} = h /\sqrt{h^2}$, and
\be
 \hat{x}^i = \hat{h}^A Z^{(i)}_A \qquad {\rm so} \qquad
\hat{x}^i\hat{x}^i = 1.
\ee
Equations \ref{eqn:gradh} and \ref{eqn:gradZ} imply
\be
 \left( d\hat{x}^i d\hat{x}^i \right)_{AB} = \left(\Delta^2 +
 h^2\right) \left(\gamma_{AB} - \hat{h}_A \hat{h}_B \right).
\ee
Note that ${\cal L}_h \hat{x}^i = 0$ so $\hat{x}^i$ is constant along
integral curves of $h$. In some open set it is therefore possible to
use $\hat{x}^i$ and the parameter along these curves as
coordinates. It is convenient to define
\be
 \mu = \frac{4}{\Delta^2 + h^2}, \qquad j = \pm
\frac{8\sqrt{h^2}}{\left(\Delta^2 + h^2\right)^2},
\ee
where the sign of $j$ will be left arbitrary. Note that $j^2 < \mu^3$.
To bring the metric to standard form, introduce polar
coordinates 
\be
 \hat{x}^1 = -\cos \phi \sin \theta, \qquad \hat{x}^2 = \sin\phi \sin
 \theta, \qquad \hat{x}^3 = \cos \theta,
\ee
and let $\psi$ be the parameter along the integral curves of $h$,
normalized so that
\be
 h =  - 4j \mu^{-5/2} \left(1-\frac{j^2}{\mu^3} \right)^{-1/2} 
 \frac{\partial}{\partial \psi} 
\ee 
The metric must take the form
\be
 ds_3^2 = \gamma_{AB} dx^A dx^B = \frac{\mu}{4} \left[ 
 \left(1 - \frac{j^2}{\mu^3} \right) \left(d\psi + {\cal A} \right)^2
 + d\theta^2 + \sin^2 \theta d\phi^2 \right],
\ee
for some locally defined $1$-form ${\cal A}$ on $H$. Equation
\ref{eqn:gradh} determines ${\cal A}$ up to a gradient, which is just
the freedom to choose the $\psi = 0$ surface. A convenient choice is
\be
 {\cal A} = \cos \theta d\phi.
\ee
which also fixes the orientation of $H$ so that $d\theta \wedge d\psi
\wedge d\phi$ is positively oriented. The metric on $H$ now takes the
form
\be
 ds_3^2 = \frac{\mu}{4} \left[ 
 \left(1 - \frac{j^2}{\mu^3} \right) \left(d\psi + \cos \theta d\phi \right)^2
 + d\theta^2 + \sin^2 \theta d\phi^2 \right],
\ee
which is the standard form of the metric on a squashed $S^3$. However,
this is a local result -- globally, $H$ could differ from a squashed
$S^3$ by discrete identifications. 

Writing $h$ as a 1-form gives
\be
 h = - j \mu^{-3/2} \left(1 - \frac{j^2}{\mu^3} \right)^{1/2} \left(
 d\psi + \cos \theta d\phi \right).
\ee

The case $h=0$ (i.e. $j=0$) is much simpler. Equation \ref{eqn:ric} establishes
that $H$ is a three dimensional compact Einstein space of positive
curvature and hence locally isometric to $S^3$ with its round
metric. Therefore the above local coordinates can also be introduced in this
case, and the metric is as above with $j=0$.

It is worth summarizing what has been shown. Local coordinates have been
introduced in a neighbourhood of the horizon ($r=0$) and explicit
expressions for the local behaviour of $\Delta$, $h_A$ and
$\gamma_{AB}$ at $r=0$ has been obtained. 
Using the expressions for the metric and field strength
(equations \ref{eqn:metric} and \ref{eqn:Fsol2}), it is now clear that
the above analysis has fully determined the local form of the
near-horizon solution. Since this local form is unique, it must agree
with that of the BMPV solution. So the above analysis proves that the
near horizon geometry of any supersymmetric black hole solution with
$\Delta >0$ on $H$ must be locally isometric to that of the BMPV
solution.

It has been proved that $H$ is locally isometric to a squashed
$S^3$ when $j \ne 0$ and a round $S^3$ when $j=0$. In the latter case,
$H$ must be globally isometric to a discrete quotient of a round $S^3$
(since $H$ is a positive Einstein 3-manifold) and in the former case, $H$ 
is presumably globally isometric to a
discrete quotient of a squashed $S^3$. The question of which quotients
are consistent with supersymmetry
can be deduced from the existence of the vector fields
$\xi^i_L$ and $h$. Whatever quotient is taken must preserve these
vector fields. 

If $j=0$ then $\xi^i_L$ generate the $SU(2)_L$ subgroup
of the $SU(2)_L \times SU(2)_R$ isometry group of $S^3$. The allowed
quotients must therefore be subgroups of $SU(2)_R$. So $H$ is
of the form $S^3/\Gamma$, where $\Gamma$ is a discrete subgroup of
$SU(2)_R$. 

If $j \ne 0$ then $\xi^i_L$ generate the $SU(2)$ subgroup, and $h$
the $U(1)$ subgroup of the $SU(2) \times U(1)$ isometry group of a
squashed $S^3$. The allowed quotients must be subgroups of $U(1)$, and
therefore cyclic groups. Hence $H$ must be a quotient of a squashed $S^3$ by a
cyclic group, i.e., $H$ is a squashed lens space. 

\subsection{Global constraints}

\label{subsec:global}

The purpose of this subsection is to prove Theorem 2, i.e., to show
that a supersymmetric black hole whose near-horizon geometry is
locally isometric to that of the BMPV solution must
actually be globally isometric to the BMPV solution. This is where
detailed knowledge of the general supersymmetric solution \cite{gauntlett:02}
of minimal $D=5$ supergravity is required.

The first step is to write the near-horizon solution in the form of
equation \ref{eqn:metricsol}. Doing so, the metric on the base space
is
\be
\label{eqn:base}
 ds_4^2 = r \Delta \left(\gamma_{AB} + \frac{1}{\Delta^2} h_A h_B
 \right) dx^A dx^B + \frac{dr^2}{r\Delta} + \frac{2}{\Delta}{dr h_A}
 dx^A.
\ee
{\it A priori}, there is no reason why this metric should be regular
at $r=0$ because the form \ref{eqn:metricsol} is only valid for $r>0$.
It has just been demonstrated that coordinates $x^A=(\psi,\theta,\phi)$
can be introduced so that the metric on $H$ is locally isometric to
that of a squashed $S^3$. In these coordinates,
\ba
 ds_4^2 &=& \frac{r \mu^{1/2}}{2} \left( 1- \frac{j^2}{\mu^3}
 \right)^{1/2} \left[ \left(d\psi + \cos \theta d\phi - j r^{-1}
\mu^{-3/2} \left(1-\frac{j^2}{\mu^3} \right)^{-1/2} dr \right)^2 +
d\theta^2 + \sin^2 \theta d\phi^2 \right] \nonumber \\
 &+& \frac{dr^2}{2r} \mu^{1/2}
\left(1-\frac{j^2}{\mu^3} \right)^{1/2} + {\cal O}(r^0) dr^2 + {\cal
O}(r) dr dx^A + {\cal O}(r^2) dx^A dx^B.
\ea
Now let
\be
 R = 2^{1/2} r^{1/2} \mu^{1/4} \left(1 - \frac{j^2}{\mu^3}
 \right)^{1/4},
\ee
and
\be
 \psi' = \psi - 2 j \mu^{-3/2}\left(1 - \frac{j^2}{\mu^3}
 \right)^{-1/2} \log R,
\ee
so
\ba
\label{eqn:flat1}
 ds_4^2 &=& dR^2 + \frac{R^2}{4} \left[ \left(d\psi' + \cos \theta d\phi
 \right)^2 + d\theta^2 + \sin^2 \theta d\phi^2 \right] \\ &+& {\cal O}(R^2)
 dR^2 + {\cal O}(R^3) dR dx'^A + {\cal O}(R^4) dx'^A dx'^B, \nonumber 
\ea
where $x'^A = ( \psi',\theta,\phi )$. 
It should be emphasized that this is only valid locally -- no
assumptions have been made about the ranges of the coordinates
$(\psi,\theta,\phi)$. However, it can be seen that the base space
metric is locally flat near $R=0$. Near $R=0$, the surfaces of constant $R$ are
positive Einstein spaces and must therefore be globally isometric to
$S^3$ with its round metric, identified under some subgroup
$\Gamma$ of its $SU(2) \times SU(2)$ isometry group. 
Generally, this implies that
there will be a conical singularity at $R=0$. The metric has to be
hyper-K\"ahler for $R>0$, so the holonomy has to be a subgroup of
$SU(2)$. Therefore the singularity at $R=0$ has to be an $A-D-E$
orbifold singularity (see \cite{aspinwall:96} for a review).\footnote{
This result also follows from the properties of $H$ deduced
in the last subsection.}

Now, $f$ approaches a constant at infinity (since $V$ is the
stationary Killing vector field), and in equation \ref{eqn:metricsol},
$\omega$ must vanish fast enough at infinity for the ADM angular
momentum to be well-defined. Asymptotic flatness then implies that the
base space has to be asymptotically Euclidean. Near $R=0$,
the metric is well-approximated by flat space with an $A-D-E$
singularity at the origin, and it is known \cite{aspinwall:96} that
such singularities can be resolved by ``blowing up'' the
singularity. This produces a complete
asymptotically Euclidean hyper-K\"ahler space. However, the only such space
is $R^4$ with its standard metric \cite{gibbons:79}. In other words,
there can't have been a singularity at $R=0$ after all!\footnote{
If one relaxes the condition of asymptotic flatness then an
$A-D-E$ singularity {\it can} occur at $R=0$. Such behaviour can by
obtained by taking appropriate quotients of the BMPV solution.}

It follows that the metric \ref{eqn:flat1} describes a 
portion of global flat space, i.e., the coordinates
$(\theta,\phi,\psi)$ must have their standard ranges $0 \le \theta \le
\pi$, $\phi \sim \phi + 2\pi$ and $\psi' \sim \psi' + 4\pi$ (which implies
$\psi \sim \psi + 4\pi$). Hence $H$ has $S^3$ topology. The base space
is globally flat, with metric 
\be
\label{eqn:flat}
 ds^2 = d\rho^2 + \frac{\rho^2}{4} \left[ \left( \sigma_R^1 \right)^2
 + \left( \sigma_R^2 \right)^2 + \left( \sigma_R^3 \right)^2 \right],
\ee
where
\be
 R = \rho + {\cal O}(\rho^3),
\ee
and
\ba
\sigma_R^1 &=& -\sin\psi' d\theta+\cos\psi' \sin\theta d\phi\nonumber \\
\sigma_R^2 &=& \cos\psi' d\theta+\sin\psi' \sin\theta d\phi\\
\sigma_R^3 &=& d\psi'+\cos\theta d\phi. \nonumber
\ea
Having established that the base space is flat, the next step is to
show that the solution must have $G^+ = 0$.
Consider equation \ref{eqn:Gplussol}. $\Delta$ is constant and
non-zero at $r=0$, which implies that $G^+$ is well-defined in a
neighbourhood of $H$. (In fact, $G^+$ vanishes when restricted to
$H$.) It is easy to see from the behaviour of $G^+$ near $r=0$
that $G^+$ is regular at the origin of the base space. Hence $G^+$ is globally
defined on $R^4$. $G^+$ must vanish at infinity in $R^4$ (because
$\omega$ has to decay fast enough for the solution to be
asymptotically flat). Furthermore $G^+$ is closed and must therefore
belong to $H_{cpt}^2(R^4,R)$, the second compactly supported cohomology
group on $R^4$. However, this group is trivial, so 
there exists a 1-form $\Gamma$ globally defined on $R^4$ 
such that $G^+ = d\Gamma$ with $\Gamma$ vanishing at infinity. Now
consider
\be
 0 = \int_{S^3_{\infty}} \Gamma \wedge G^+ = \int_{R^4} G^+ \wedge G^+
 = \int_{R^4} d^4 x \left( G^+ \right)^2,
\ee
where the first integral is taken over the three sphere at infinity,
and the final equality follows from the self-duality of $G^+$. Hence
$G^+$ must vanish everywhere on the base space and therefore
everywhere in the spacetime.

The vanishing of $G^+$ implies (equation \ref{eqn:delf}) that $f^{-1}$
is harmonic on the base space. Near the origin,
\be
 f^{-1} = \frac{1}{r \Delta} = \frac{\mu}{\rho^2} + g,
\ee
where $g$ is ${\cal O}(\rho^0)$ near $\rho=0$. As mentioned above, for
a single black hole, it can be assumed that $f$ is positive everywhere
outside the black hole, and hence $f^{-1}$ must be finite for $\rho >
0$. Furthermore, $f$ must approach a constant at infinity. This
implies that $g$ is regular and bounded on $R^4$. But $g$ must be
harmonic (as $f^{-1}$ is) and hence constant. By rescaling the
coordinates, $g$ can be set to $1$, so
\be
\label{eqn:fsol}
 f^{-1} = 1 + \frac{\mu}{\rho^2}.
\ee
The final step is to prove that $\omega$ is uniquely
determined. Topologically, it is clear that $\omega$ can be globally defined
defined on $R^4 - \{0\}$ by equation \ref{eqn:omegadef}. Now
\ba
 - f^{-2} V &=& du - \frac{dr}{r^2 \Delta^2} - \frac{h}{r \Delta^2} \nonumber \\
  &=&  dt + \frac{j}{4r} \mu^{-1/2} \left( 1 - \frac{j^2}{\mu^3}
 \right)^{-1/2} \sigma_R^3 + {\cal O}\left(r^{-1} \right) dr + {\cal
   O}\left( r^0 \right) dx^A,
\ea 
where
\be
 t = u + \frac{\mu}{4r}
\ee
is a time coordinate defined for $r>0$.
Hence, up to a gradient (which can be absorbed into $t$), $\omega$ is
given by
\ba
\label{eqn:omegaorigin}
 \omega &=&  \frac{j}{4r} \mu^{-1/2} \left( 1 - \frac{j^2}{\mu^3}
 \right)^{-1/2} \sigma_R^3 + {\cal O}\left(r^{-1} \right) dr + {\cal
   O}\left( r^0 \right) dx^A, \nonumber \\
   &=& \frac{j}{2\rho^2} \sigma_R^3 + {\cal O}(\rho^{-1})d\rho + {\cal
   O} (\rho^0) dx^A.
\ea
For $\rho > 0$, $\omega$ has to satisfy the following criteria. First,
the vanishing of $G^+$ implies that $d\omega$ has to be anti-self-dual
with respect to the metric on the base space. Secondly, asymptotic
flatness requires $\omega = {\cal O}(\rho^{-3})$ as $\rho \rightarrow
\infty$. Finally, as $\rho \rightarrow 0$, $\omega$ must be given by
equation \ref{eqn:omegaorigin}. To prove uniqueness of $\omega$,
assume that there were two solutions $\omega_1$ and $\omega_2$
satisfying these criteria. Let $\tilde{\omega} = \omega_1 -
\omega_2$. Hence $\tilde{\omega}$ is ${\cal O}(\rho^{-3})$ as $\rho
 \rightarrow \infty$, and, near $\rho = 0$,
\be
 \tilde{\omega} =   {\cal O}(\rho^{-1})d\rho + {\cal
   O} (\rho^0) dx^A.
\ee
Therefore $\tilde{\omega}$ can be written, on $R^4 - \{0\}$, as
\be
 \tilde{\omega} = \frac{\alpha}{\rho} d\rho + \nu,
\ee
where $\nu \equiv \nu_A dx^A$ is a ($\rho$-dependent) $1$-form defined on $S^3$. The
quantities $\alpha$ and $\nu$ are well-behaved as $\rho \rightarrow 0$.
Anti-self duality of $d\tilde{\omega}$ reduces to
\be
 \hat{d} \alpha = *_3 \hat{d} {\nu} + \rho \partial_\rho \nu,
\ee
where $\hat{d}$ and $*_3$ are now defined on the round unit $S^3$. This
equation  implies that $\alpha$ becomes a harmonic function on $S^3$
as $\rho \rightarrow 0$. However, the only harmonic functions on $S^3$
are constant, so $\alpha$ must become constant as $\rho \rightarrow
0$, i.e., $\hat{d} \alpha \rightarrow 0$ as $\rho \rightarrow 0$.
Therefore $\hat{d}{\nu} \rightarrow 0$ as  $\rho \rightarrow 0$. However this implies
\be
 0 = \lim_{\epsilon \rightarrow 0} \int_{\rho = \epsilon} \tilde{\omega} \wedge
 d\tilde{\omega} = \int_{R^4-\{0\}} d\tilde{\omega} \wedge d\tilde{\omega} =
 - \int_{R^4-\{0\}} d^4 x \left(d \tilde{\omega} \right)^2,
\ee
where the surface term at infinity vanishes because of the boundary
conditions on $\tilde{\omega}$, and the final equality is a
consequence of anti-self-duality. It follows then, that
$\tilde{\omega}$ is closed and hence $\tilde{\omega} = d \lambda$ for
some function $\lambda$ defined for $\rho > 0$. Therefore, $\omega_1$
and $\omega_2$ differ at most by a gradient. Hence the general solution for $\omega$ 
must agree with the BMPV solution up to a gradient:
\be
 \omega = \frac{j}{2\rho^2} \sigma_R^3 + d\lambda.
\ee
Finally, this gradient can be absorbed into the time coordinate $t$,
and then
\be
 V = -f^2 \left(dt + \frac{j}{2\rho^2} \sigma_R^3 \right)
\ee
for $r>0$. Since the base space is flat, and $f$ is given by equation
\ref{eqn:fsol}, the solution is identical to the BMPV solution for
$r>0$. 

\subsection{Discussion}

The above results constitute the first examples of uniqueness theorems for
supersymmetric black holes. Theorem 1 was proved by a local analysis
of the constraints imposed by supersymmetry in a neighbourhood of the
horizon. It was shown that the near-horizon geometry is completely
determined. The proof of Theorem 2 was based on the 
general form for all supersymmetric solutions determined in 
\cite{gauntlett:02}. Asymptotic flatness and the boundary conditions
obtained from the near-horizon geometry select a unique solution from
this class.

Note that Theorem 1 determines the near-horizon
geometry of {\it any} (i.e. not necessarily asymptotically flat)
supersymmetric solution that admits a compact Killing horizon preserved by $V$. The
near-horizon geometry of such a solution has to be locally isometric
to flat space, to $AdS_3 \times S^2$, or the near-horizon geometry of
the BMPV solution (of which $AdS_2 \times S^3$ is a special
case). Furthermore, in each case, the allowed possibilities for the
spatial geometry of the event horizon (i.e. of $H$) have been
determined. In the flat case, $H$ must be $T^3$ with its flat metric and
in the $AdS_3 \times S^2$ case, $H$ must be $S^1 \times S^2$ with the
usual metric. In the case of a BMPV near-horizon geometry there are
more possibilities. If $j \ne 0$ then $H$ must be a squashed lens
space. If $j=0$ then the near-horizon BMPV geometry reduces to $AdS_2
\times S^3$, and in this case $H$ must be a quotient of a round $S^3$ by a
discrete subgroup of $SU(2)$.

Even if a spacetime has a non-compact Killing horizon, it is often possible to
make identifications to render the horizon compact. For example,
$AdS_3 \times S^2$ arises as the near-horizon geometry of a magnetic black
string wrapped around a compact Kaluza-Klein direction, with momentum
around this direction. A supersymmetric spacetimes admitting a Killing horizon 
for which the analysis of this paper does {\it not} determine the
near-horizon geometry would have to satisfy one of two 
criteria. Either the horizon would not be preserved by $V$, or the
horizon would be non-compact and could not be rendered compact by
identifications without breaking supersymmetry. The former case is not
of much physical interest since one is usually interested in event
horizons, which {\it must} be preserved by all Killing vector fields.

It would be interesting to see whether the above
method could be extended to prove uniqueness theorems for other
supergravity theories. For example, proving uniqueness of
supersymmetric black holes in minimal $N=2$, $D=4$ supergravity
amounts to proving the long-standing conjecture \cite{hartle:72} that
the only black holes in the Israel-Wilson-Per\'jes (IWP) class of solutions
are the Mujumdar-Papapetrou multi-black hole solutions. This is
because all supersymmetric solutions of this theory are known
\cite{tod:83}, and fall into a timelike and a null class, as for the
minimal $D=5$ theory. The IWP solutions constitute the timelike
class. It seems very likely that this conjecture could be proved
easily using the methods of this paper, i.e., first constructing bosonic objects
from the super-covariantly constant spinor, using these to determine
the form of the near horizon geometry (presumably either flat space or
$AdS_2 \times S^2$), and then showing that this information together with
asymptotic flatness determines a unique member of the IWP
family.

Of more physical interest would be the extension of the above results
to more complicated supergravity theories, for example the maximally 
supersymmetric theories in $D=4,5$. 
It seems rather unlikely that the general supersymmetric
solution of these theories could be obtained using the methods of
\cite{tod:83,gauntlett:02}, so a complete uniqueness proof is probably not
possible using the methods of this paper. However, it might be
possible to determine all possible near-horizon geometries of
solutions with compact Killing horizons (one would start with the
metric \ref{eqn:metric} and assume that $\Delta$, $h_A$ and
$\gamma_{AB}$ are independent of $r$, since this is what happens in
the near-horizon limit). This information might lead
to an understanding (in classical supergravity)
of why supersymmetric rotating black hole solutions only
seem to exist in $D=5$. Finally, it might be possible to use the
methods of the present paper to classify possible near-horizon
geometries of supersymmetric solutions of $D=10,11$ supergravity
theories.

\bigskip
\bigskip

\centerline{\bf Acknowledgments} 

\medskip

I would like to thank Fay Dowker, Steven Gubser, Gary Horowitz, 
Chris Hull, Juan
Maldacena, Robert Wald, Daniel Waldram, Toby Wiseman and Edward Witten
for discussions, and Roberto Emparan, Jerome Gauntlett, Gary
Gibbons, and especially James Sparks for discussions and comments on the
manuscript. I am grateful to Amanda Peet for pointing out a typo in
the first version of this paper. This work was supported by PPARC.

\renewcommand{\theequation}{A.\arabic{equation}}

\setcounter{equation}{0}

\section*{Appendix}

The purpose of this appendix is to examine the possibility of
supersymmetric black holes with near-horizon geometry $AdS_3 \times
S^2$. Some constraints on the form of the base space of such a
solution will be obtained. 
It is best read after subsection \ref{subsec:global} because it
relies on results developed there.

Since $\Delta$ vanishes at $r=0$, but is not identically zero (this
would correspond to the null class) then by analyticity it must be 
possible to write
\be
\label{eqn:pdef}
 \Delta = r^p \tilde{\Delta},
\ee
where $p$ is a positive integer, $\tilde{\Delta}$ is not
identically zero on $H$, and $\tilde{\Delta}>0$ for $r>0$.
The Maxwell equation \ref{eqn:delf} then
gives
\be
\label{eqn:specialcase}
 \nabla^2 \tilde{\Delta} - (2p-1) h^A \nabla_A \tilde{\Delta} = -
 p(p-1) \tilde{\Delta} h^2 + {\cal O}(r).
\ee
Integrating this equation over $H$ (using $\nabla_A h^A = 0$)
implies
\be
 p(p-1) \tilde{\Delta} h^2 = 0 \qquad \mbox{on $H$}
\ee
Since $h^2 \ne 0$ on $H$ for the solution \ref{eqn:ads3}, it follows
that this solution must have $p=1$. Substituting this back into equation
\ref{eqn:specialcase}, multiplying by $\tilde{\Delta}$ and integrating
over $H$ gives
\be
 \tilde{\Delta} = \tilde{\Delta}_0 \qquad \mbox{on $H$},
\ee
where $\tilde{\Delta}_0$ is a positive constant.
Next, from subsection \ref{subsec:special}, $h$ can be written as
\be
 h = -\frac{dR}{R} + r h_1 + r^2 h_2,
\ee
where $h_1$ is independent of $r$ and $h_2$ is smooth at $r=0$.
Equation \ref{eqn:dh} implies
\be
 \hat{d} (Rh_1) = 0.
\ee
Hence there is some function $\lambda$ defined locally on $H$ such
that
\be
 h_1 = \frac{L}{R} d\lambda.
\ee 
The base space metric can be calculated from equation
\ref{eqn:base}:
\be
 ds_4^2 = \tilde{\Delta}^{-1} \left(\frac{r}{R} \right)^2 \left[
 d\left( L\lambda - \frac{R}{r} \right) + Rrh_2 \right]^2 + r^2
 \tilde{\Delta} \gamma_{AB} dx^A dx^B.
\ee
Define a new coordinate $\rho$ by
\be
 r = \frac{\rho}{X},
\ee
where
\be
 X = 1 + \frac{L \lambda \rho}{R}.
\ee
The base space metric is then
\be
 ds_4^2 = X^{-2} \left\{ \tilde{\Delta}^{-1} \left[ \frac{d\rho}{\rho}
- \frac{dR}{R} + \frac{\rho^2 h_2}{X} \right]^2 + L^2 \rho^2
\tilde{\Delta} \left[ \frac{dR^2}{R^2} + d\Omega^2 + {\cal O}(\rho)
 dx^A dx^B \right]  \right\},
\ee
where the metric on $H$ deduced in subsection \ref{subsec:special} has
been used and $x^A = ( R,\theta,\phi )$. Completing the square on $dR/R$ gives
\ba
 ds_4^2 &=& X^{-2} \left\{ \tilde{\Delta}^{-1} Y \left[ \frac{dR}{R} -
 Y^{-1} \left( \frac{d\rho}{\rho} + \frac{ \rho^2 h_2}{X} \right)
 \right]^2 + \tilde{\Delta} Y^{-1} L^2 \rho^2 \left[\frac{d\rho}{\rho}
 + \frac{ \rho^2 h_2}{X} \right]^2 \right. \nonumber \\
 &+& \left. L^2 \rho^2 \tilde{\Delta} d\Omega^2 +
 {\cal O}(\rho^3)dx^A dx^B \right\},
\ea 
where
\be
 Y \equiv 1 + L^2 \rho^2 \tilde{\Delta}^2.
\ee
Hence
\ba
 ds_4^2 &=& \left( 1 + {\cal O}(\rho) \right) \left\{
 \tilde{\Delta}_0^{-1} \left[ \frac{dR}{R} - \frac{d\rho}{\rho} +
 {\cal O}(\rho) d\rho + {\cal O}(\rho^2) dx^A \right]^2
 \right. \nonumber \\ &+& \left. L^2 \tilde{\Delta}_0 \left( d\rho^2 +
 \rho^2 d\Omega^2 \right) + {\cal O}(\rho) d\rho^2 + {\cal O}(\rho^3)
 d\rho dx^A + {\cal O}(\rho^3) dx^A dx^B \right\}
\ea
Now define
\be
 x^4 = \tilde{\Delta}_0^{-1/2} \log \frac{R}{\rho}, \qquad
 \tilde{\rho} = L \tilde{\Delta}_0^{1/2} \rho.
\ee
Then
\ba
\label{eqn:flatbase}
 ds_4^2 &=& d\trho^2 + \trho^2 d\Omega^2 + (dx^4)^2
 + {\cal O}(\trho) d\trho^2 + {\cal O}(\trho) d\trho dx^4 +
 {\cal O}(\trho^3) d\trho dx'^A \nonumber \\  &+&{\cal O}(\trho)
 (dx^4)^2 + {\cal O}(\trho^2) dx^4 dx'^A + {\cal O}(\trho^3) dx'^A dx'^B,
\ea
where $x'^A = (x^4,\theta,\phi )$. The
identifications inherited from $H$ imply that $\theta$ and $\phi$
parametrize a two-sphere and $x^4$ is periodically
identified. Superficially, the base space is therefore regular at
$\tilde{\rho}=0$. However, more careful inspection reveals that not
all of the correction terms above are necessarily smooth at $\trho=0$
so this conclusion may be incorrect. Asymptotic flatness requires the
base space to be asymptotically Euclidean. Hence if the base space is
regular then it must be global $R^4$ with its flat metric. 

It is not clear whether supersymmetric black holes with near-horizon geometry $AdS_3
\times S^2$ actually exist. Of course $AdS_3 \times S^2$ does arise as the near horizon geometry
of a black {\it string}. Such strings exist in both the null class and the timelike
class. To get such a string from the timelike class, one can take the base
space to be $R^4$ parametrized as in \ref{eqn:flatbase} (neglecting
the corrections) and then follow the method of section 3.7 of
\cite{gauntlett:02} 
(with $H=1$, $\chi_i = \omega_i = 0$, and point sources for the harmonic
functions).

\end{document}